# Monte Carlo analysis of the contributions of long-lived positronium to the spectra of positron-impact-induced secondary electrons measured using an annihilation-gamma-triggered time-of-flight spectrometer

S. Lotfimarangloo[1*], V.A. Chirayath[1], S. Mukherjee[2], H. Akafzade[1], A. J. Fairchild[1], R. W. Gladen[1], A. R. Koymen[1], and A. H. Weiss[1#]

[1]Department of Physics, University of Texas at Arlington, Texas, USA-76019
[2] Radiochemistry Division, Bhabha Atomic Research Center, Mumbai, India-400085

**Abstract**

Magnetic bottle Time-of-Flight (ToF) spectrometers can measure the energy spectra of all electrons emitted into a 2π sr solid angle simultaneously, greatly reducing data collection time. When the detection of the annihilation gamma (γ) and the detection of the electron (e) are used as timing signals for ToF spectrometers, the e-γ time difference spectra (e-γ TDS) are reflective of the positron-induced electron energy distributions provided the times between the impact of the positrons and the emission of the annihilation gammas are short compared to the flight times of the electrons. This is typically the case since positrons have short lifetime in solids (~100 – 500 ps) compared to the flight times of the secondary electrons ($10^2$ ns to $10^3$ ns). However, if the positron leaves the surface as a positronium atom (a bound electron-positron state), the annihilation gamma photons can be appreciably delayed due to the longer ortho-positronium (o-Ps) lifetime. This can result in an e-γ TDS having an exponential tail with a decay constant related to the o-Ps lifetime. Here, we present an analysis of the e-γ TDS using a Monte Carlo model which estimates the spectral contributions resulting from o-Ps annihilations. By removing the contributions from the delayed gamma signal, the energy spectrum of Positron Impact-Induced Secondary electrons (PIISE) can be isolated. Furthermore, our analysis allows an estimation of the intensity of the exponential tail in the e-γ TDS providing a method to measure the fraction of positrons that form Ps at solid surfaces without relying on assumed 100% Ps emitting surfaces for calibration.

[*]sima.lotfimarangloo@mavs.uta.edu, mailto:[#]weiss@uta.edu

## 1. Introduction

The interaction of low-energy positrons with solid surfaces results in multiple signals that can be used to probe the electronic or chemical structure of surfaces [1]. Low-energy positron bombardment may result in the emission of electrons from surfaces through multiple mechanisms, including Auger mediated positron sticking (AMPS) [2,3], positron-impact-induced secondary electron emission (PIISE) [4-7] and positron annihilation-induced Auger decay processes [8,9]. PIISE occurs when positrons have sufficient kinetic energy to knock one or more electrons out of the solid. This process, which has a direct analog to electron-induced secondary electron emission (SE), becomes dominant once the kinetic energy of the positron incident on the sample is a few eV above the minimum energy required for secondary electron emission given by:

$$KE_{min,e^+} = \varphi^- - \varphi^+ \qquad (1).$$

Here $\varphi^-$ is the electron work function of the analyzer, and $\varphi^+$ is the positron work function of the sample.

Measurements of the positron-induced secondary yield, δ, defined as the number of electrons emitted per incident positron, and the energy distribution of electrons emitted as a result of positron impact provide insight into fundamental aspects of particle solid interactions. These measurements are important for understanding positron detection using electron multipliers and to investigate the (undesirable) emission of secondary electrons from the walls of particle accelerators [10-12]. As pointed out by Weiss *et al.* [5,7], and later by Overton and Coleman [13], the energy spectrum of positron-induced secondary electrons can provide important insights into the energy spectrum of electron-induced secondary electrons. This is because positrons have the same mass as electrons and similar collisional cross sections. Thus, the energy loss mechanisms for positrons in solids are similar to those of electrons. A major advantage of using positrons to study secondary electron emission is that positrons are distinguishable from electrons due to the positron's positive charge. This enables the differentiation between backscattered incident particles and true secondary electrons that may prove considerably difficult using electron-stimulated secondary electron emission. The ability to distinguish between backscattered incident particles and true secondary electrons may be highly beneficial in cases where features in secondary electron energy distribution are used to investigate the unoccupied levels of 2D materials grown on interacting substrates [14].

Previous measurements of positron-impact-induced secondary electron emission have been made with electrostatic positron beams [4-7, 15, 16]. More recently, magnetic bottle ToF spectrometers have been used to measure positron-induced electron emission [17, 18]. The magnetic bottle ToF technique is highly efficient due to the parallel collection of all electrons within a wide range of energies which are emitted into a 2π sr solid angle, greatly reducing the data collection time. In ToF measurements, the energies of electrons emitted from the sample are determined from the time it takes for the electrons to traverse a known distance. For measurements of the energies of positron-induced electrons from surfaces it is necessary to have a timing signal that corresponds to the time the electron leaves the surface and another timing signal that corresponds to the arrival of the electron after it has traversed a known distance [17, 18]. The ToF method was first implemented for measurements of positron-induced electron emission by Suzuki et al. [19] using a pulsed positron beam. The electron ToF was determined from the time difference between a signal corresponding to the arrival of the positron pulse and a signal from the detection of the electron at a channel plate.

A different scheme, first employed for the measurement of secondary electron energies by the University of Texas at Arlington (UTA) group [17, 18, 20], utilizes the signal from the detection of the annihilation gamma as a timing signal. The energies of the positron-induced-electrons are determined by measuring the differences between the detection of the electrons and the detection of the annihilation gamma rays to obtain an electron-gamma time difference spectrum (e-γ TDS). The advantage of this second method is that it does not require a pulsed beam allowing for simpler transport and the possibility of measurements employing lower beam energies. For example, in the ToF system of Suzuki et al. [19], the time compression techniques used to efficiently obtain high resolution timing result in an inherent energy spread of several eV. This energy spread combined with transport considerations of the bunched beam places a lower limit on the incident beam energy of pulsed beam ToF systems of the order of 10's of eV [19]. In contrast, the annihilation gamma timing scheme permits measurements with incident beam energies of order 1 eV, a full order of magnitude lower than for the pulsed positron beam systems [17, 18, 20].

The interpretation of the e-γ TDS is simplest for events in which the time difference between the emission of the measured electron and the emission of the annihilation gamma ray is short in comparison to the ToF of the electrons (as is the case for positron annihilation-induced Auger electrons [21]). For these events, the time difference between the detection of the electron and the detection of the annihilation gamma corresponds directly to the electron's ToF. In particular, the time between the emission of the secondary electron and the emission of the annihilation gamma can be neglected when the positron annihilates in a bulk state or surface state since the bulk state lifetime is ~100 ps and the surface state lifetime is ~500 ps [22]. Similarly, when the positron annihilates, after leaving the sample, as a positronium atom (Ps) in the singlet state, which has a mean lifetime of 125 ps, [22] the time difference between the emission of the escaping electron and the emission of the annihilation gamma can be neglected. However, if the incident positron is reemitted as Ps in the long-lived triplet state ($^3S_1$), which has a mean lifetime of ~ 142ns in vacuum, the time difference between the detection of the annihilation gamma and the detection of the electron can be significantly shorter than the actual ToF. In fact, it is even possible for the electron to be detected before the detection of the annihilation gamma ray leading to a clearly unphysical, negative ToF.

For the surfaces and incident positron beam energies presented in this study, it is estimated that the fraction of incident positrons reemitted as triplet Ps is in the range of ~ 20% to 50%. Therefore, in order to obtain the true ToF spectra of positron-impact-induced secondary electrons, these spectral contributions due to the delayed gamma associated with triplet Ps must be accounted for. In this paper, we discuss an analysis method that employs a Monte-Carlo simulation of the emission, transport, and annihilation of the Ps atoms in our ToF spectrometer which models the time distribution of the annihilation gamma resulting from o-Ps annihilation. The time distribution of the annihilation gamma from o-Ps annihilations is used in an iterative fitting scheme that decomposes the e-γ TDS into two parts. One part is associated with events in which the annihilation gamma photons were produced by the annihilation of positrons in short lifetime spin states, and the second is associated with the annihilation of positrons in long lifetime spin states. We show that this decomposition can be used to extract the true electron ToF spectrum from the e-γ TDS even in the case where a significant number of events detected involve the annihilation gammas from long-lived Ps.

In addition, our modeling of the e-γ TDS data allows an estimate of the fraction of incident positrons that are reemitted into the vacuum as Ps. Prior experiments have effectively used this method to accurately measure the lifetime of o-Ps in vacuum [23] and to perform depth-resolved lifetime measurements without a pulsed positron beam [24]. These experiments were optimized for measuring the lifetime and hence, had the secondary electron detector close to the sample for reducing the spread in the time distribution of secondary electrons reaching the detector. Additionally, the sample geometry, incident positron energy, sample chamber, and annihilation gamma detection geometry were optimized to strongly confine the emitted Ps spatially in order to maximize the detection efficiency of the three-gamma annihilation of the o-Ps atoms. In contrast, the ToF spectrometer used in our experiments has the electron detector ~ 1 m from the sample which means the long-lived o-Ps is not strongly spatially confined and can potentially travel outside the field of view of the gamma detector prior to annihilation, thus reducing the detection efficiency. Using the Monte Carlo simulation of the Ps emission and transport from the sample, we obtain a detector efficiency correction factor that accounts for the field of view of our detector allowing an estimate of the Ps fraction as a function of incident positron kinetic energy. The measurement of Ps fraction as a function of incident positron energy provides significant insight into the dynamics of the positron-surface interaction [25] and is important for surface spectroscopic techniques like Ps ToF spectroscopy [26, 27]. The accurate detection of Ps fraction is also important for experiments that aim to produce high intensity Ps for creation of Bose-Einstein condensate of Ps [28, 29] and for surface diffraction experiments involving high intensity Ps beam [30].

## 2. Experiment

The measurements presented in this paper were obtained using the low-energy positron beam equipped with a 1-meter flight path ToF spectrometer described previously [17] and shown schematically in Fig. 1. The ToF spectrometer employs an axial magnetic field to guide the positrons from a tungsten foil moderator to the sample and to guide the electrons emitted from the sample back to a microchannel plate (MCP). A pair of **E×B** plates is used to drift the incoming positrons below the MCP and a second pair is used to drift the positrons back onto the beam axis. This second pair of **E×B** plates also drifts the electrons from the sample up into the MCP where they are detected. The ToF tube provides a flight path of uniform potential for the electrons travelling from the sample to the MCP. The pulse resulting from the detection of an electron by a MCP is used as a start signal for a Time-to-Amplitude Converter (TAC) and the pulse resulting from the detection of an annihilation gamma ray by a BaF$_2$ scintillator is delayed and used as the stop signal. The pulses from the TAC are pulse-height-analyzed and, using a PC based multi-channel analyzer, a histogram of the measured time difference is generated yielding the time difference spectra whose analysis is discussed below. The sample is polycrystalline Cu that was cleaned using argon ion sputtering. The positrons were transported through the ToF tube at an energy of less than 1 eV and then accelerated to the indicated incident positron beam energies by biasing the sample negatively with respect to the ToF tube. The experiments were performed at room temperature unless otherwise indicated.
We define a time difference variable:

$$\tau = t_\gamma - t_e + C \quad (2),$$

where $t_e$ is the time of detection of a positron-induced electron by the MCP and $t_\gamma$ is the time of detection of the gamma resulting from the annihilation of that positron and $C$ is a constant that is chosen for convenience in plotting. Our experimental e-γ TDS is given by histogram $N(\tau)\Delta\tau$, which is defined as the number of events in a bin of width $\Delta\tau$ having a time difference τ.

5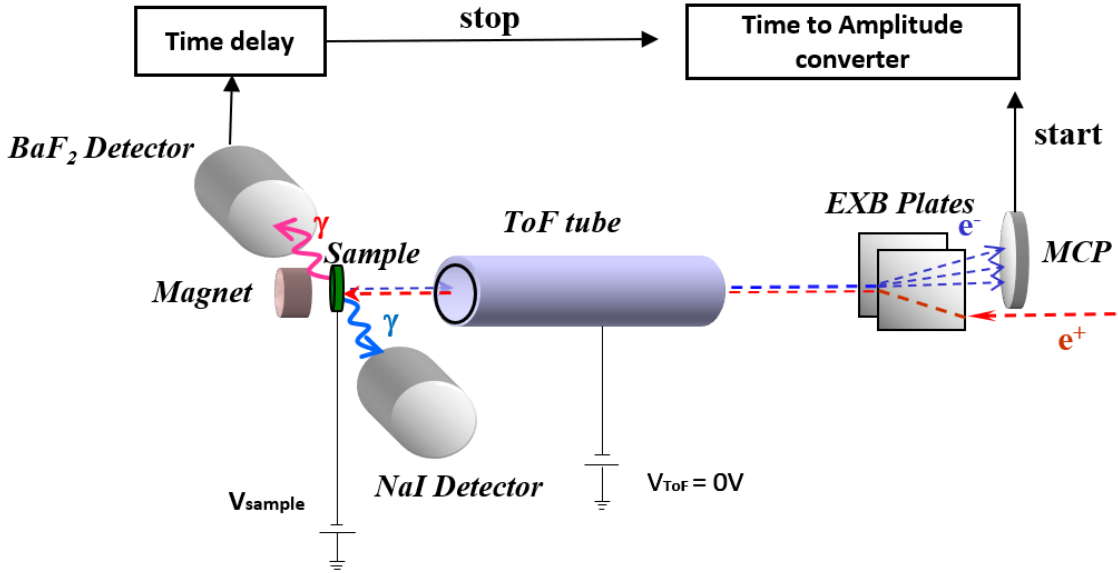

Fig.1. Schematic of the UTA 1 m ToF spectrometer. The time difference ($\tau$) between the detection of the 511 keV γ, by the BaF$_2$ detector, and the detection of the secondary electron, by the MCP, is used to determine the kinetic energies of the emitted secondary electrons.

## 3. Results and Discussion
### 3.1. Exponential Tail in the time difference spectra

Fig. 2 shows the time difference spectrum, obtained from a polycrystalline Cu sample biased at -60 V, in which the incident positrons have a maximum kinetic energy of 61 eV. In this case, the constant $C$ was chosen to make '$\tau = 0$' correspond to an event in which the annihilation gamma and the electron were emitted simultaneously, and the electron was emitted from the surface of the negatively biased sample with kinetic energy (KE) = 0 eV. The solid red line indicates a value, of $\tau_{max}$, corresponding to an event in which the gamma ray and electron were emitted simultaneously, and the electron was emitted normal to the surface with a KE of 61 eV, the maximum KE with which positron-impact induced secondary electrons can leave the surface. The exponentially decreasing tail observed for $\tau > \tau_{max}$ corresponds to events in which the annihilation gamma is associated with long-lived triplet Ps. The purple line is a fit to the exponential tail from which we obtain an o-Ps annihilation lifetime of ~120 ns, which is ~ 15% lower than the o-Ps lifetime of 142 ns in vacuum. This indicates that some of the o-Ps are quenched due to annihilations with the wall or due to the flight of o-Ps away from the field of view of the gamma detector. We will discuss these points in the modeling of exponential tail in section 3 below.

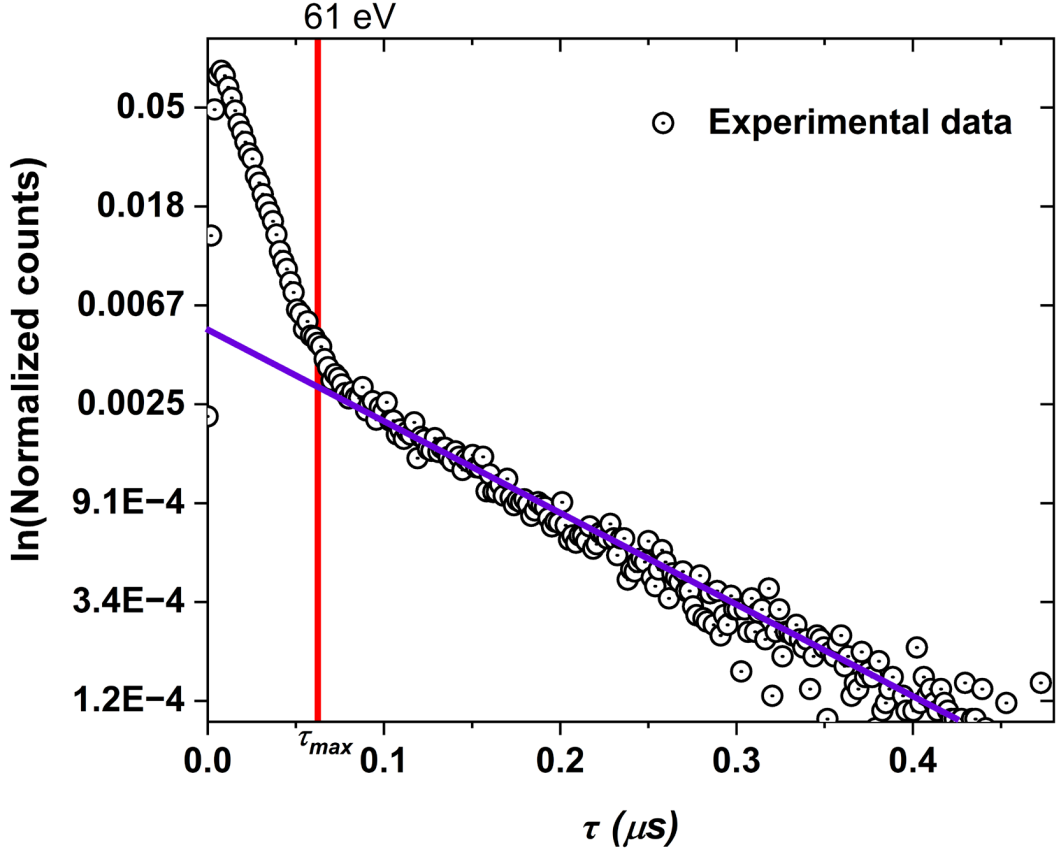

Fig. 2. Time difference spectrum obtained from a polycrystalline Cu sample biased at – 60 V corresponding to a maximum incident positron energy of 61 eV. A constant $C$ has been added to the time difference to make the 0 value of $\tau$ correspond to events in which the electron is leaving the surface of the negatively biased sample at 0 eV. The red line corresponds to events in which the electron was emitted normal to the surface with the maximum KE (61 eV) with which positron-impact induced secondary electrons can leave the surface. The purple line is a fit to the exponential tail to obtain the o-Ps lifetime.

### 3.2. Modeling of the Experimental Time Difference Spectrum $N(\tau)d\tau$

In modeling the experimental time difference histogram, we assume that it can be approximated in terms of an appropriately scaled and normalized density function $N(\tau)d\tau$ representing the probability of observing an event with a time difference $\tau$ in an interval of $d\tau$. We note that $\tau$ depends on the time of detection of the electron and the time of detection of a corresponding gamma ray. We define two functions: $N_e(t_e)dt_e$, the probability of detecting an electron at a time $t_e$ within a time interval $dt_e$, and $N_\gamma(t_\gamma)dt_\gamma$, the probability of detecting a gamma ray in a time $t_\gamma$ within a time interval $dt_\gamma$. We take $t_e = 0$ to correspond to the time of emission of the positron-induced electrons. Therefore, $N_e(t_e)dt_e$ corresponds to the distribution of the positron-induced electron flight times. We take $t_\gamma = 0$ to correspond to the time at which the positron impinges upon the sample which we assume to be the same time as the emission of the electron from the surface. In doing so we neglect time differences of the order of hundred picoseconds or less corresponding to the time between the entry of positron and (i) the emission of secondary electrons, (ii) the emission of Ps into the vacuum, (iii) the falling of the positron into the surface state and (iv) the thermalization of positron into a Bloch wave or its localization in a trap in the bulk.

With the definitions above, the joint probability of detecting both a gamma ray and an electron may be written as:

$$P(t_e, t_\gamma)dt_e dt_\gamma = N_e(t_e)N_\gamma(t_\gamma)dt_e dt_\gamma \quad (3).$$

Performing a change of variables using equation (2) and integrating over all $t_e$, we find the probability of observing an event with the value $\tau$ within an interval $d\tau$ is given by:

$$N(\tau - C)d\tau = \left[\int_{-\infty}^{\infty} N_\gamma(t_e + \tau - C)N_e(t_e)dt_e\right]d\tau \quad (4),$$

which is a form of the cross-correlation integral.

To represent $N_e(t_e)$ in a way which is consistent with how we have displayed ToF data in our previous publications we define an associated distribution $N'_e$ such that:

$$N'_e(t_{neg})dt_{neg} = -N_e(t_e)dt_e \quad (5).$$

We define $t_{neg} = C - t_e$ with $C$ appropriately chosen as described above. substituting this new function into equation (4) and applying the appropriate limits of integration we get:

$$N(\tau)d\tau = \left[\int_{-\infty}^{\infty} N_\gamma(\tau - t_{neg})N'_e(t_{neg})dt_{neg}\right]d\tau \quad (6),$$

which is in the form of a convolution integral.

For the purposes of measuring the energy distribution, $N_e(E)dE$, of PIISE emitted from the sample, we use an iterative deconvolution scheme, described in detail below, to extract the ToF distribution function $N'_e(t_{neg})$ from the experimental data, $N(\tau)d\tau$. We then use a flight time-to-energy transformation (obtained using a calibration method described in [17]) to transform the distribution $N'_e(t_{neg})dt_{neg}$ into the distribution $N_e(E)dE$.

To accomplish this effective deconvolution, we must first obtain a model for the function $N_\gamma(t_\gamma)dt_\gamma$. We use Monte Carlo modeling of the positron annihilation from the triplet state and singlet state to obtain an approximation for $N_\gamma(t_\gamma)dt_\gamma$ which is combined with equation (6) to extract $N'_e(t_{neg})dt_{neg}$ from the measured data using an iterative fitting scheme described in detail in a later section. Our method is equivalent to performing an iterative deconvolution of instrumentally broadened spectra. In our experiment this process is facilitated by two factors:

1. The lifetime of positrons annihilating in the singlet state is at least 1000 times shorter than other relevant times (including the time-of-flight of the secondary electrons and the lifetime of positrons annihilating in a triplet Ps state).

Consequently, we can approximate the time distribution $N_\gamma(t_\gamma)$ as follows:

$$N_\gamma(t_\gamma) = \alpha N_{\gamma,singlet}(t_\gamma) + \beta N_{\gamma,triplet}(t_\gamma) \cong \alpha\delta(t_\gamma) + \beta N_{\gamma,triplet}(t_\gamma) \quad (7),$$

where $N_{\gamma,singlet}(t_\gamma)$ is the probability distribution for detecting a gamma resulting from the short lifetime singlet annihilations and $N_{\gamma,triplet}(t_\gamma)$ is the probability distribution for detecting a gamma resulting from the annihilation of a positron originally in a triplet spin state of Ps, and $\alpha$ and $\beta$ are constants to be determined. Here, it is assumed that $N_{\gamma,singlet}(t_\gamma)$ can be replaced by a delta function as it is narrow (100-500 ps) compared to other time scales in the experiment (100 ns - 1µs). Substituting equation (7) into (6) results in:

$$N(\tau)d\tau = \alpha\,N'_e(\tau)d\tau + \beta\left[\int_{-\infty}^{\infty} N_{\gamma,triplet}(\tau - t_{neg})N'_e(t_{neg})dt_{neg}\right]d\tau \qquad (8),$$

from which it is possible to extract the "true" ToF spectrum of secondary electrons $N'_e(\tau)$ as well as the coefficients $\alpha$ and $\beta$.

2. There is a sharp cut off in the distribution of electron flight times: $N'_e(t_{neg}) = 0$ for $t_{neg} > t_{neg,max}$ where $t_{neg,max}$ is the value of $t_{neg}$ corresponding to an electron traveling through the ToF tube with the maximum energy allowed by energy conservation:

$$KE_{max,e} = -2eV_{sample} + KE_{e^+} + \varphi^+ - \varphi^- \qquad (9).$$

Here $KE_{max,e}$ is the maximum kinetic energy of electrons traversing the ToF tube, $e$ is the magnitude of charge of the electron, $V_{sample}$ is the potential on the sample relative to the ToF tube, $KE_{e^+}$ is the energy of the positron as it traverses the ToF tube, $\varphi^+$ is the positron work function of the sample, and $\varphi^-$ is the electron work function of the analyzer. Because of these two factors, the region of measured times $\tau$ that correspond to long-lived Ps can be clearly identified in the measured time difference spectra as a nearly exponential tail that begins at a known value ($\tau_{max}$ when expressed in terms of $\tau$ or $t_{neg,max}$ when expressed in terms of $t_{neg}$). The ability to identify a clear exponential tail facilitates the application of the iterative fitting procedure described later.

### 3.3. Monte Carlo modeling of the distribution of gamma detection times for triplet Ps $\left(N_{\gamma,triplet}(t_\gamma)\right)$

As noted above, the exponential tail observed in the experimental data is indicative of the presence of long-lived Ps. A simple linear fit to the log of this tail, purple line in Fig. 2, yields a lifetime of ~120 ns. Monte Carlo simulations of the Ps emission, transport, and annihilation described below shows that this shortened lifetime can be accounted for because of two mechanisms resulting from the motion of the Ps atoms away from the sample surface:

(i). The collisions of outward moving Ps with a wall resulting in the reduction of the mean o-Ps lifetime through rapid two gamma pick-off annihilation.

(ii). The decrease in the efficiency of detection of the annihilation gammas from Ps as it moves out of the scintillator detector's line of sight.

To model the effects of these two mechanisms, Monte-Carlo techniques were used to generate an ensemble of model Ps atoms with a selected pseudorandom distribution of initial angles, initial speeds, and pre-assigned times of self-annihilation. The Ps atoms were generated one at a time and followed in time and position through the simulated chamber until they annihilated. If the preassigned time of self-annihilation was sufficiently large the Ps atom trajectory could intercept the chamber wall where it could either annihilate at the wall through pickoff or bounce with an assigned reflection probability. After the Ps atom annihilated either in the triplet state or via two gamma pick-off annihilation on the wall, its contribution to $N_{\gamma,triplet}(t_\gamma)$ is weighted according to the efficiency of detection of the annihilation event as determined from the solid angle subtended by the detector at position of Ps annihilation and the number of gamma rays produced during the annihilation event. Details of the various steps are given below. The algorithm is consistent with a similar method employed by Mariazzi et al. [28] and Rienäcker et al [31], where they used Monte Carlo simulations of Ps transport and annihilation inside a vacuum chamber to obtain the annihilation gamma energy spectrum. One key difference of our model is our assumption that some of the Ps atoms that reach the chamber walls can reflect. Mariazzi et al. [26], assumed in their model that the Ps ejected from the nanochanneled silicon membranes annihilated as soon as it hit the chamber walls. This difference is due to the assumed energy distributions of the modeled Ps atoms of the two methods. Ps atoms in our model are assumed to be formed by fully thermalized positrons that reach the surface and are ejected with an energy of a few eV (~2 eV). Mariazzi et al. [26], assumed that the Ps atoms ejected from the silicon membranes have undergone large number of reflections inside the nanochannels and are ejected with only few tens of milli eV energy which then annihilate without reflection from the chamber walls [23].

For generating the ensemble of initial model Ps atoms, we assumed that the Ps atoms were emitted from the center of the sample with a pseudorandom angular distribution given as:

$$N(\theta)d\theta = 2\sin\theta\cos\theta\, d\theta \quad (10),$$

where $\theta$ is measured with respect to the surface normal. This distribution is more forward directed than the isotropic distribution $\sin\theta\, d\theta$. An isotropic distribution would result in a higher density of Ps emitted at larger $\theta$ values that would produce a peak in the e-$\gamma$ TDS beyond the primary secondary electron peak. This second peak would correspond to Ps annihilations after hitting the walls of the chamber. The absence of such peak in our experimental time difference spectra of secondary electrons with different incident positron energies points to a more forward directed Ps emission from the surface.

The initial speeds of the Ps atoms were chosen from a pseudorandom Gaussian distribution with a mean energy of 1.5 eV and a standard deviation of 0.4 eV. This is consistent with the energy spectrum of thermalized Ps emitted from a Cu surface [32]. The maximum energy of the Ps that can be emitted from the sample when a fully thermalized positron reaches the surface is:

$$KE_{max,Ps} = 6.8\, eV - (\varphi^+ + \varphi^-) \quad (11).$$

Here 6.8 eV is the binding energy of Ps. For Cu, $\varphi^+$ is small (close to zero) but negative [2] and $\varphi^-$ is ~ 4.5 eV [33]. Hence, the maximum energy of Ps would be ~ 2 eV for the case where Ps is formed using a fully thermalized positron at the surface. With increasing incident positron energy (from 10 eV – 900 eV), the probability for epithermal positrons to reach the surface increases and thus, there is a greater probability for the emission of higher energy Ps. The present model does not consider variation in Ps energy distribution because of the formation of epithermal Ps formation.

At birth, each Ps atom is preassigned a self-annihilation time that corresponds to the time it would annihilate in an infinitely large vacuum chamber (i.e., in the absence of wall quenching) given by the distribution:

$$N(t_\gamma)dt_\gamma = N_0 \exp\left(-\frac{t_\gamma}{\tau_{Ps}}\right) dt_\gamma \quad (12),$$

where $\tau_{Ps}$ = 142 ns, the o-Ps lifetime in vacuum. After the Ps atom is assigned its initial emission angles ($\theta$, $\phi$), speed, and time of self-annihilation, its position of annihilation is determined as a function of time. If the value of the preassigned annihilation time was large enough to allow the Ps atom to reach the chamber wall, a pseudorandom generator is used to choose, based on the assumed reflection probability, from two possible outcomes: annihilation at the time of impact via pickoff (resulting in two gammas) or reflection from the chamber wall without annihilation. The reflection probability was treated as a variable parameter. For simplicity, the reflections were assumed to be elastic and specular. The code allowed for following the Ps through multiple reflections if required. If the Ps atom did not annihilate with the chamber wall through pickoff until its preassigned time of self-annihilation, it will self-annihilate in between the walls with the emission of three gammas. The position and time of annihilation of each Ps atom was recorded. A histogram of the times of detection of the annihilation gammas was then produced by weighting the contribution of each annihilation event according to the calculated detection probability.

Fig. 3 (a) shows the coordinate axis, origin, and the definition of angles used for determining the detection probability. The origin of the axis is at the center of the sample surface. The center of the BaF$_2$ scintillation detector is at $(0, -R, 0)$, where $R \sim 2.54$ cm is the outer radius of the ToF tube. In this coordinate system, the positions of the Ps atoms when they annihilate, either via pick-off or self-annihilation, are shown in Fig. 3(b). Only events corresponding to $z = 0$ are shown for simplicity. The simulation results show that most of the Ps atoms annihilate within 1cm of the sample. The second most preferred annihilation position is on the walls. We found the best agreement with the measured time difference spectrum when the Ps atoms had an 80% probability to reflect from the walls. Once the annihilation positions are known, the geometrical detection efficiency is determined from the solid angle subtended by the detector at the point of annihilation as shown in Fig. 4 (a). Even though the BaF$_2$ scintillation detector is a 2-inch x 2-inch cylindrical crystal, we have considered only the solid angle subtended by the front circular face of the detector at the position of annihilation. The BaF$_2$ scintillation detector has ~ 8 mm of Pb and steel shielding around it resulting in ~70% reduction in the intensity of 511keV gamma reaching the detector from the sides and hence, treating the detector as circular disk is justified. The shielding was introduced to reduce gamma background from the source and to prevent detection of 511 keV gamma from annihilations at the **E** x **B** region. The small solid angle $d\Omega$ subtended by the detector area element $d\vec{A}$ at the annihilation position ($\vec{r}$) is given by:

$$d\Omega = \frac{d\vec{A}\cdot\vec{r'}}{r'^3} = \frac{dx\, dz\, (y_0+R)}{((x-x_0)^2+(z-z_0)^2+(y_0+R)^2)^{\frac{3}{2}}} \qquad (13).$$

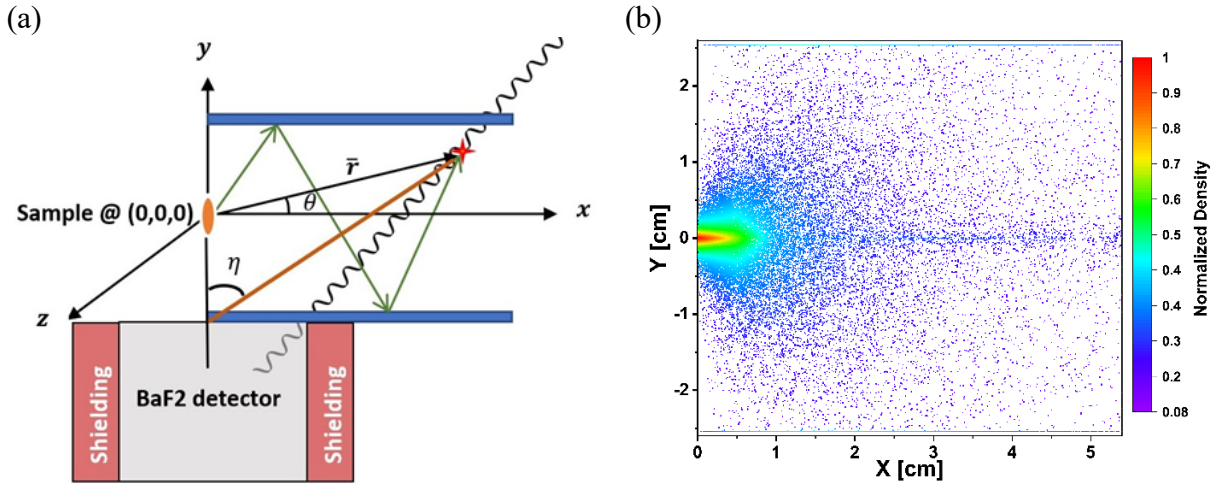

(a)      (b)

Fig. 3 (a). Schematic of the model used to simulate the Ps emission, annihilation, and detection using the Monte Carlo method. The green arrows show a Ps atom ejected from the sample surface and reflecting off the walls before annihilating at $\bar{r}$. The annihilation position is also designated using angle $\eta$, defined with respect to the normal to the detector surface. (b) 2D Heat map showing the distribution of the annihilation positions for Ps atoms which were emitted from the sample with a mean energy of 1.5 eV (standard deviation of 0.4 eV) from (0, 0, 0) with an 80% chance of reflecting from the walls of the chamber. Only those events where z = 0 is shown for clarity.

Integrating equation (13) over the circular detector face yielded the total solid angle expressed in terms of the angle $\eta$ for different distances from the center of the detector (Fig. 4(b)). The calculated solid angles are consistent with previous analytical and numerical results [34]. The ratio of the calculated solid angle to $2\pi$ gives the geometric efficiency of the BaF$_2$ scintillator in detecting the 511 keV gamma rays that originate from a singlet annihilation event. For annihilation from the triplet state, the solid angle corresponding to the detection of at least one of the three annihilation gamma rays varies from a minimum of $4\pi/3$ to a maximum of $4\pi/2$ depending on the angle between the three gamma rays. From previous measurements of the number of detected gamma photons and the Ps fractions from a copper substrate before and after removal of a single layer graphene [35], we estimate that the solid angle corresponding to detection of one of the three o-Ps annihilation gamma is $4\pi/2.89$. Therefore, the ratio of the calculated solid angle to $4\pi/2.89$, gives the geometric efficiency of the BaF$_2$ scintillator in detecting triplet annihilation events. Based on this geometric detection efficiency, a particular Ps atom annihilation event is included or not in the time distribution histogram of the annihilation gammas (Fig. 5(a)). The time distribution histogram of the detected annihilation events, $N_{\gamma,triplet}(t_\gamma)$, is shown in Fig. 5(b).

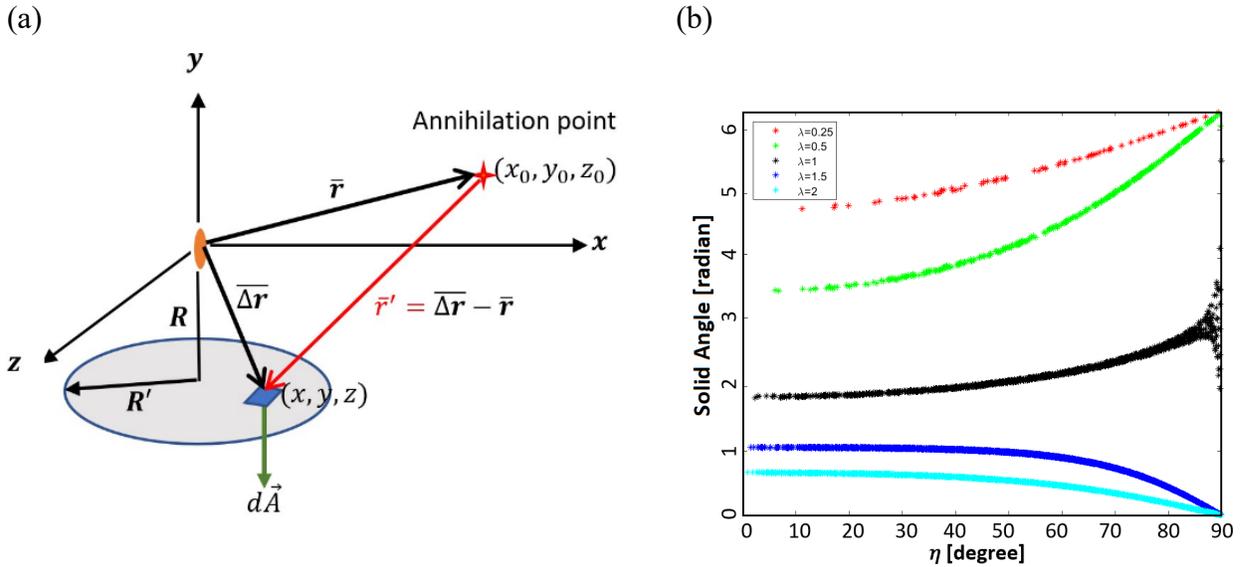

Fig. 4 (a): Solid angle $d\Omega$ subtended by $dA$ at the point of annihilation calculated using equation (13) based on the vectors shown. The total solid angle is obtained by integrating over the circular face of the detector of radius $R'$. (b) Solid angle calculated as a function of $\eta$ for various values of $\lambda$. Here $\lambda$ is the ratio of the annihilation point distance from the center of the detector's circular surface and $R'$.

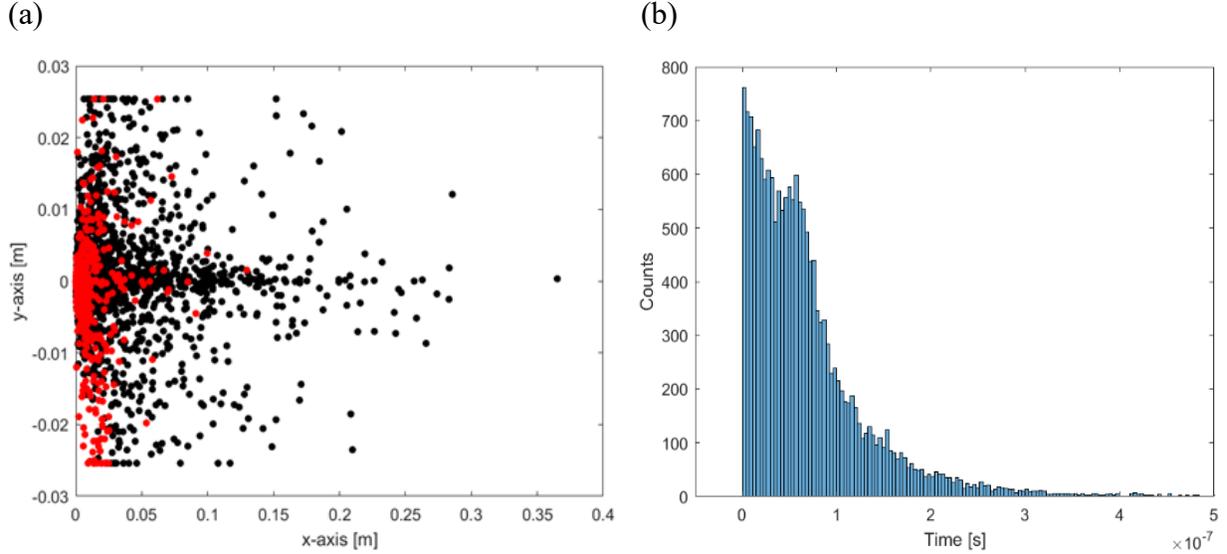

Fig. 5 (a) All Ps atom annihilations events (black) compared to the annihilation events that are detected by the BaF$_2$ scintillation detector (red) after considering the geometric efficiency of detector. (b) Time distribution histogram of gamma annihilation events detected by the BaF$_2$ detector for o-Ps emitted from surface.

### 3.4. Iteration scheme to extract the distribution of electron flight times $(N'_e(\tau))$.

The first estimate of the electron flight time distribution function $N'_{1,e}(\tau)$ is obtained by truncating the experimental spectrum $N(\tau)d\tau$ at the maximum possible energy of the secondary electron corresponding to $\tau_{max}$ (refer to Fig. 2) and subtracting a straight line from $\tau = 0$ to $\tau = \tau_{max}$, see Fig. 6(a). The resulting estimate, $N'_{1,e}(\tau)$, is area normalized (Fig. 6(b)) and convoluted with the area normalized annihilation gamma time distribution function (Fig. 6 (c)) obtained using the Monte Carlo simulation described above. The result of the convolution is shown in Fig. 6(d) which is the first estimate of the e-γ TDS of secondary electrons that were tagged only with the o-Ps annihilation events. Then, the spectrum shown in Fig. 6(d) is used in a one-parameter least squares fit to the exponential tail (for $\tau > \tau_{max}$) of the experimental spectrum (Fig. 6 (a)) to obtain $\beta$. The result is shown in Fig. 6(e). Then the fit is subtracted from the experimental spectrum to obtain the next estimate of the "true" electron flight time distribution function $N'_{2,e}(\tau)$. The new estimate is area normalized (Fig. 6(f)) and the iteration (steps 6(d) to 6(f)) is continued until $\frac{|\beta_i - \beta_{i-1}|}{\beta_{i-1}} < 5 \times 10^{-5}$. In the case of e-γ TDS from clean Cu obtained with 60 eV positrons, the convergence was obtained by the fourth iteration. This rapid convergence even when the Ps fraction was ~ 50% provides strong evidence that our results do not strongly depend on the details of the modeling of long-lived Ps related events. In fact, these events contribute roughly only 1% for time differences relevant to measurements of secondary electron spectra. Fig. 7 shows a comparison between the experimental spectrum and the final estimate of the the electron flight time distribution function without any contribution from o-Ps annihilation events produced by the iterative fitting scheme (PIISE-red), and the convolution of the PIISE with the model annihilation gamma time distribution (blue).

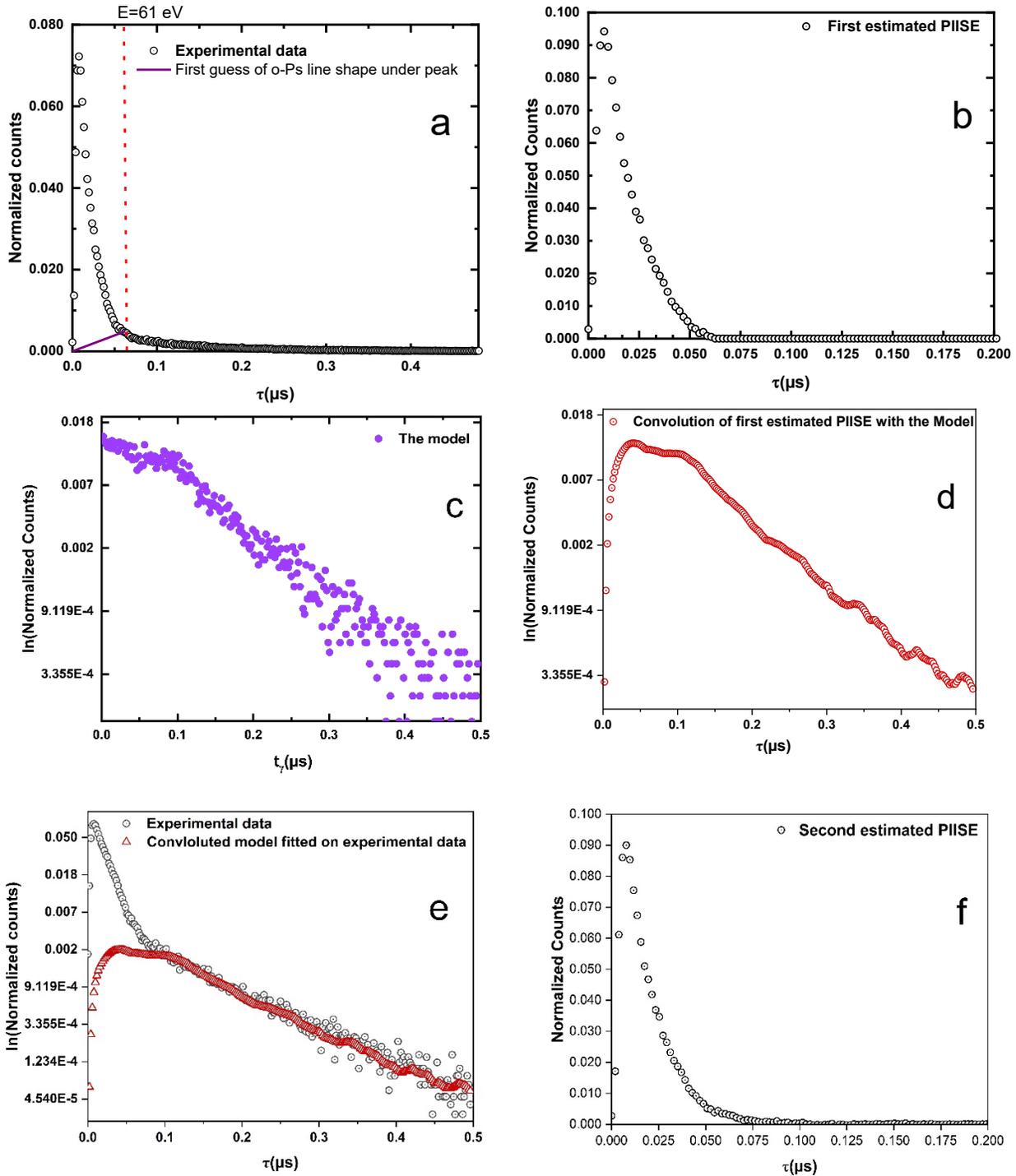

Fig.6. Secondary electron spectra (PIISE) emitted from a polycrystalline Cu surface following the implantation of 60 eV positrons and the steps applied on the time spectrum to get the true secondary electrons spectra. (a) Truncating the spectra at maximum possible energy (shown by dotted red line) and subtracting a straight line (solid purple line) from τ = 0 to $τ_{max}$ (b) Resulting first estimate of true PIISE spectra (c) time distribution of gamma detected as a result of o-Ps annihilations simulated using Monte Carlo methods (d) Convolution of the first estimate of PIISE spectra and o-Ps gamma time distribution giving an estimate of the e-γ TDS corresponding to PIISE events tagged by o-Ps annihilations (e) Fitting the exponential tail of the experimental spectra with the estimated e-γ TDS corresponding to PIISE events tagged by o-Ps annihilations. To show the fit to the exponential tail clearly, the plot is shown in log scale (f) The second estimate of the true PIISE spectra.

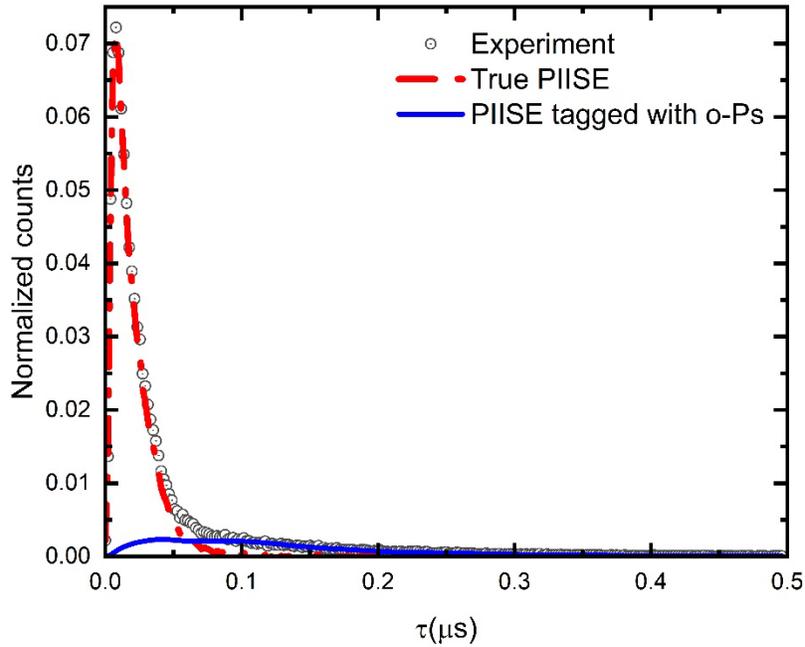

Fig. 7: The true electron flight time distribution function without any contribution from secondaries tagged with o-Ps annihilation (PIISE- red) and the convolution of the PIISE with the model annihilation gamma time distribution function (blue).

**3.5. Application of the iterative fitting method to find the energy distribution of positron-induced secondary electrons from Cu**

We applied the methods described in the previous section to find the e-γ TDS of PIISE from clean Cu measured with positron energies ranging from 10 eV to 900 eV. These time difference spectra were then converted to energy spectra using a transformation function obtained through the calibration procedure described in [8]. The energy spectra deduced from the e-γ TDS, after removing the o-Ps contribution (a few representative spectra are shown in Fig.8), are consistent with measurements of the energy of PIISE [5,7].

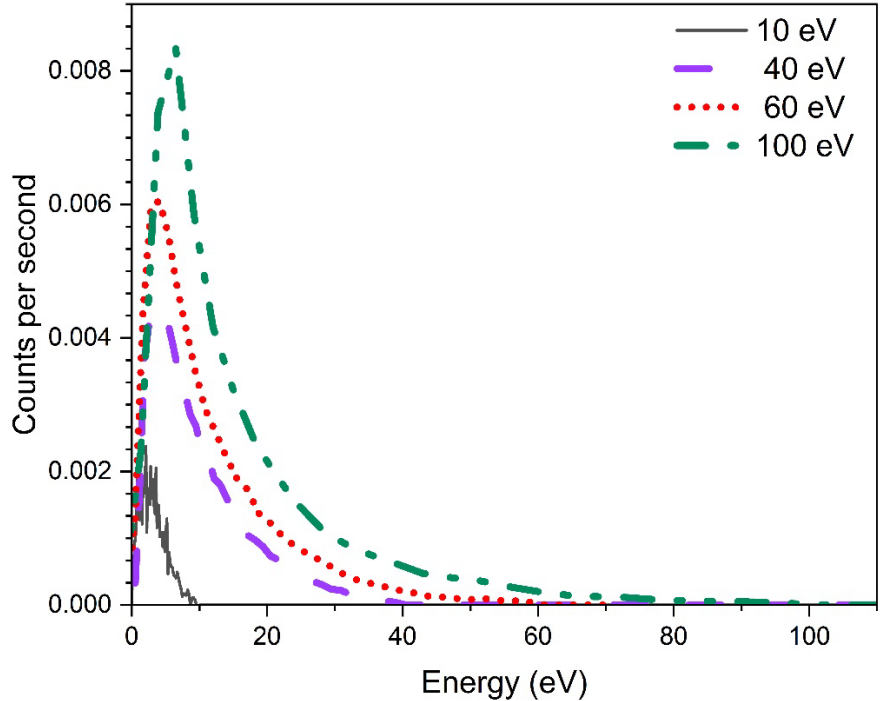

Fig.8. Positron-induced secondary electron energy spectrum for different positron energies determined from the measured e-γ TDS spectra converted to energy. The spectral component due to the long-lived o-Ps has been removed using the method described above. The maximum electron energies are now consistent with what is expected from the incident kinetic energy of the positrons shown in the legends.

**3.6. Test of the applicability of the iterative fitting method to measure the secondary electron energy spectrum in cases where the Ps fraction approaches 100%**

As noted earlier, even with ~ 50% Ps formation from clean Cu surface, the contribution of the secondary electrons that are tagged with the o-Ps annihilations were not large enough to produce significant modifications to the PIISE. However, to demonstrate the general applicability of the iterative scheme to obtain electron spectra that are untagged with delayed o-Ps annihilations, we used our method to analyze an e-γ TDS where ~ 88% of the electron spectrum was associated with Ps annihilations. In their pioneering experiments, Mukherjee et al. showed that energy released during the transition of the positron from a scattering state to a bound surface state, called the Auger mediated positron sticking (AMPS) process, can be large enough to cause the emission of an electron even when the incident positron energy is lower than the threshold for PIISE emission [2]. Fig. 9 shows their measurement of the e-γ TDS for Cu at a maximum incident positron energy of 3.3 eV, which is below the minimum energy required for PIISE (equation (1)), but above the minimum energy required for AMPS. Moreover, detailed modeling of the AMPS line shape at similar incident positron energies for Cu has shown that the e-γ TDS spectra at 3.3 eV does not contain any electrons that are tagged with Ps atoms [3]. Fig. 9 also shows the e-γ TDS for Cu at a maximum incident positron energy of 1.25 eV, where the only energetically possible electron emission process is the Auger relaxation of annihilation-induced holes by surface trapped positrons annihilating with 3p and 3s core electrons of Cu. The Auger relaxation of these core holes results in the emission of 60 eV $M_{2,3}VV$ and 108 eV $M_1VV$ Auger electrons. By subtracting e-γ TDS spectra collected at 1.25 eV from that collected at 3.3 eV (after normalizing the $M_{2,3}VV$ peaks), we obtain the spectrum of electrons emitted via AMPS alone (inset of Fig. 9(a)). When the experiment was repeated at 993 K, it was observed that the Auger peaks reduced in intensity by ~88% and the AMPS spectrum widened (Fig. 9(b)). This was explained by Mukherjee et al. [2] using the model put forward by S. Chu et al. [36] that considered a two-step process for the thermal desorption of Ps atoms from metal surfaces: (i) The positron sticks to the surface emitting AMPS electrons and (ii) majority of the surface trapped positrons escape from the surface potential well as Ps atoms reducing the intensity of Auger electron peaks. Since, the majority of AMPS electrons are now tagged with Ps annihilation, the resulting AMPS spectra gets a prominent tail due to the tagging of the AMPS electron with the delayed gamma rays. The result is electron intensities at energies higher than is energetically possible through the AMPS process. To obtain the spectrum of only AMPS electrons emitted from the hot surface shown in Fig. 9(c), we subtracted the room temperature Auger spectra collected at 1.25 eV after normalizing the Cu $M_{2,3}VV$ Auger peak, shown in Fig. 9(b). We applied our iterative scheme (with 0.128 eV Ps and a reduced reflection coefficient of 50%) to extract the "true AMPS spectra" from the distorted AMPS spectra collected at 993 K. The success and general applicability of our method is demonstrated by the likeness of the AMPS spectra extracted using our method from the 933 K data to the AMPS spectra collected at room temperature (Fig. 9(d)). The estimate of the AMPS spectrum extracted was smoothed using an adjacent averaging algorithm to overcome the low counts for measurements taken at 993K. We used the "true AMPS spectra" in our iterative fitting scheme to obtain $α$ and $β$ which gave a Ps fraction of 100±10 % which is consistent with the drastic reduction in the Auger intensity.

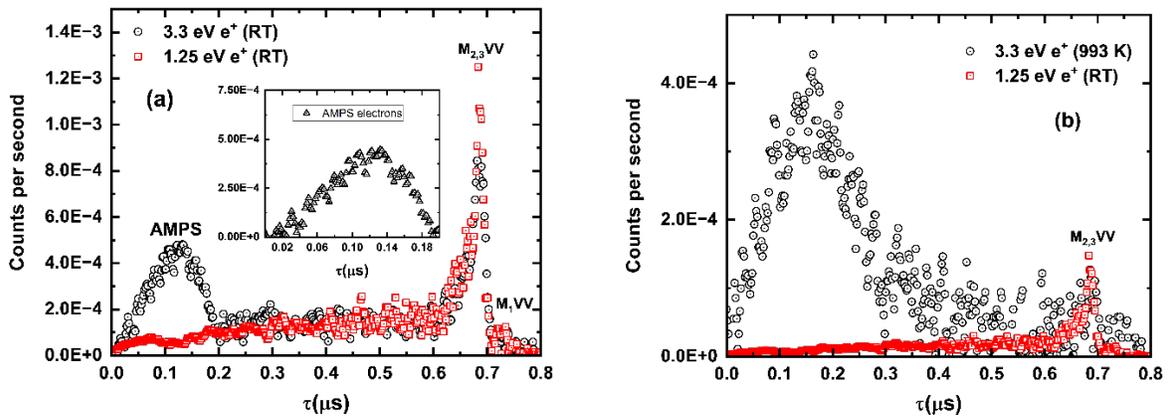

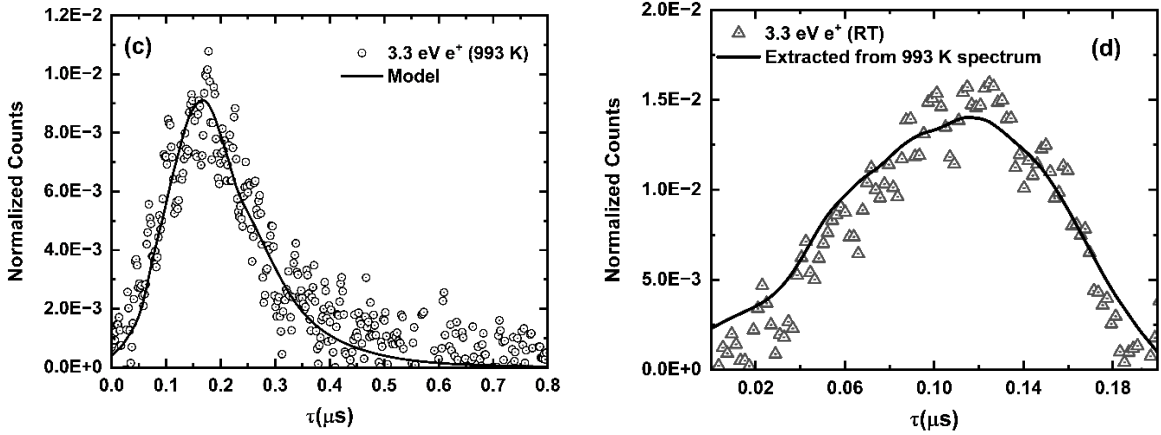

Fig.9. (a) e-γ TDS spectrum from clean Cu obtained at room temperature (RT) with maximum incident positron energies of 3.3 eV (black circles) and 1.25 eV (red squares). For the e-γ TDS with 1.25 eV positrons all electrons are associated with two gamma annihilations from the surface state resulting in sharp Auger peaks. With an incident positron energy of 3.3 eV there are also electrons emitted through the AMPS process. The inset shows the spectrum of AMPS electrons alone after removing the spectra of Auger electrons as described in the text. (b) The e-γ TDS collected with a maximum incident positron energy of 3.3 eV at 993 K is compared to the $M_{2,3}VV$ normalized e-γ TDS collected with a maximum incident positron energy of 1.25 eV. The drastic reduction (~88%) in the intensity of the Auger peak shows that majority of the positrons escape from the surface state as Ps and hence, majority of the AMPS electrons are tagged with annihilations from the Ps state. The electrons emitted through the AMPS process is obtained by subtracting the e-γ TDS with 1.25 eV positrons after normalizing their $M_{2,3}VV$ Auger peak. (c) AMPS-induced electron spectrum at 993 K (black circles) with the simulated spectrum (black line) using the Monte Carlo model. (d) Comparison of the AMPS spectrum measured at room temperature with the "true AMPS spectrum" obtained through the Monte Carlo based iterative deconvolution scheme.

### 3.7 Application of the iterative fitting method to obtain an independent estimate of the Ps fraction

The iterative fitting method gives the fraction of PIISE or AMPS electrons that were tagged with singlet annihilations (*α*) and with the delayed gamma following o-Ps annihilations (*β*). Since these parameters are dependent on the fraction of positrons that form Ps and on the fraction that are detected, it is possible to estimate the absolute Ps fraction from the iterative scheme after obtaining the *α* and *β* values. Let $f_{Ps}$ be the fraction of positrons which form Ps, *ρ* the fraction that are reflected, $\delta(KE_e, KE_{e^+})$ the number of secondary (or AMPS) electrons of energy $KE_e$ emitted following the impact of a positron with kinetic energy $KE_{e^+}$, $\varepsilon(KE_e)$ the transport efficiency of electron with kinetic energy $KE_e$, $G_{sample}$ the geometric detection efficiency of two gamma singlet annihilations at or near the sample, and $G_{o-Ps}$ be the geometric detection efficiency of o-Ps annihilations. Then the number of electrons with energy $KE_e$ associated with singlet annihilations is:

$$X = (1 - \rho)(1 - f_{Ps})\delta(KE_e, KE_{e^+})\varepsilon(KE_e)\ G_{sample} + 0.25(1 - \rho)f_{Ps}\delta(KE_e, KE_{e^+})\varepsilon(KE_e)G_{sample} \qquad (14),$$

and the number of electrons associated with o-Ps annihilations is:

$$Y = 0.75(1 - \rho)f_{Ps}\delta(KE_e, KE_{e^+})\varepsilon(KE_e)G_{o-Ps} \qquad (15),$$

with $\alpha$ and $\beta$ given by $\frac{\int X d(KE_e)}{\int (X+Y)d(KE_e)}$ and $\frac{\int Y d(KE_e)}{\int (X+Y)d(KE_e)}$ respectively. Our modeling shows that $G_{sample}$ is 0.29 and $G_{o-Ps}$ is 0.216. $G_{o-Ps}$ considers both the two 511 KeV γ photons produced during wall annihilations and 2.8 γ photons on average produced during vacuum annihilations. Fig. 10 is a comparison between $f_{Ps}$ as a function of incident positron kinetic energy obtained from the iterative scheme and $f_{Ps}$ obtained from gamma spectroscopy measured independently using a NaI scintillation detector (see Fig. 1 for arrangement). $f_{Ps}$ from gamma spectroscopy was obtained from the valley to peak ratio (R-parameter) as described in [37] where it is assumed that the gamma spectrum measured from Ge at 1000K corresponds to 100% Ps [24]. Recently, however, Reinäcker et al. [29] contested this assumption using experimental gamma spectroscopy and Monte Carlo simulations and showed that at 1000 K only 72.8% of positrons that fall on Ge form Ps [29]. We also calculated $f_{Ps}$ based on the results in [29]. The $f_{Ps}$ obtained from analyzing the secondary electron spectrum using the iterative scheme lies closer to those calculated from the NaI gamma data that assumed ~73% positronium formation from Ge at 1000 K.

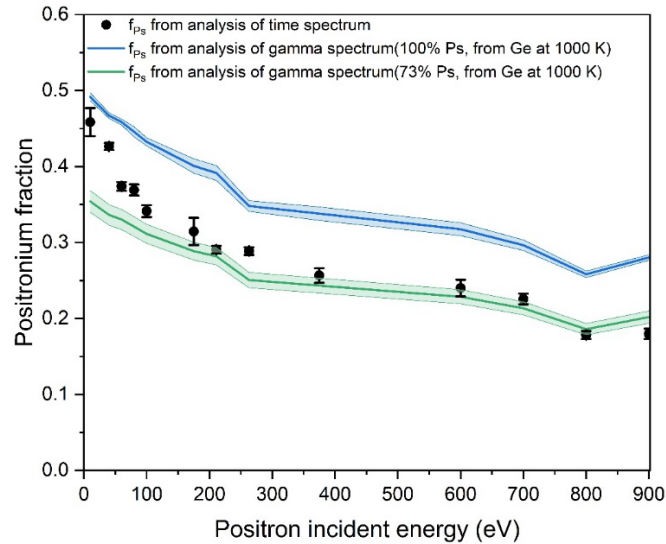

Fig. 10. Variation in the fraction of positrons that form Ps at a Cu surface for different incident positron energies (10eV to 900eV). The black circles are $f_{Ps}$ obtained from the iterative scheme. $f_{Ps}$ calculated from gamma spectroscopy assuming 100% Ps formation at 1000 K from a Ge surface is shown in blue solid lines [37]. $f_{Ps}$ calculated assuming $(72 \pm 0.4)\%$ Ps formation from a Ge surface at 1000 K is shown by green solid lines [31].

**Conclusion:**
We have presented a method to remove positronium-induced artifacts from the energy spectra of positron-induced electrons measured using a time-of-flight spectrometer in which the timing signals are obtained from the detection of the electron and the detection of an annihilation-induced gamma ray. We used a Monte Carlo model to simulate the spectral contributions resulting from o-Ps annihilations. Using a method equivalent to an iterative deconvolution scheme, we were able to decompose our measured spectra into two parts: one corresponding to the detection of gamma rays from the annihilation of positrons in a combination of short lifetime states (bulk, surface, and singlet Ps) and another corresponding to the detection of gamma rays from the annihilation of positrons in long lifetime triplet Ps states. Our results demonstrate that the method can extract the true energy spectrum even when a large fraction of the incident positrons leave the surface as long-lived o-positronium. Furthermore, the rapid convergence of the iterative scheme provides compelling evidence that the e-$\gamma$ ToF spectroscopy may be used with confidence even in cases where the Ps fraction is large. Lastly, our method provides a means to estimate the fraction of positrons that form Ps at solid surfaces without the need for calibration using assumed 100% Ps emitting surfaces.

**Acknowledgements:**
AHW, VAC and ARK acknowledge the support of NSF Grant No. CHE – 2204230. AHW and ARK acknowledge the support of NSF Grant No. NSF-DMR-1338130 & NSF-DMR1508719. AHW acknowledges the support of Welch Foundation (Y-1968-20180324).